\documentclass[a4paper,11pt]{article}
\pdfoutput=1 

\usepackage{jcappub} 
\usepackage[english]{babel}
\pdfoutput=1
\usepackage[utf8x]{inputenc}
\usepackage{epsfig,latexsym,cancel,amssymb,amsmath,xcolor}
\usepackage{hyperref}         
\usepackage{graphicx}
\usepackage{psfrag,rotating}
\usepackage{lscape}
\usepackage{enumerate}
\usepackage{amsfonts}
\usepackage{caption}

\renewcommand{\thetable}{$\arabic{table}$}

\newcommand{\mpl}{M_\mathrm{Pl}}

\def\beq{\begin{equation}}
\def\eeq{\end{equation}}
\def\bea{\begin{eqnarray}}
\def\eea{\end{eqnarray}}

\def\ln#1{\mathrm{ln}\left(#1\right)}

\title{The paradigm of warm quintessential inflation and spontaneous baryogenesis }

\author[a]{Soumen Basak,}
\author[b]{Sukannya Bhattacharya,}
\author[c]{Mayukh R. Gangopadhyay,}
\author[d]{Nur Jaman,}
\author[e]{Raghavan Rangarajan,}
\author[d]{and M. Sami} 
\affiliation[a]{School of Physics, Indian Institute of Science Education and Research Thiruvananthapuram, Maruthamala PO, Vithura, Thiruvananthapuram 695551, Kerala, India}
\affiliation[b]{Centre for Strings, Gravitation and Cosmology, Department of Physics, Indian Institute of Technology Madras, Chennai 600036, India}
\affiliation[c]{Centre For Theoretical Physics, Jamia Millia Islamia, New Delhi 110025, India.}
\affiliation[d]{Centre for Cosmology and Science Popularization(CCSP), SGT University, Gurugram 12006, Delhi-NCR, India.\\
Center for Theoretical Physics, Eurasian National University,
Astana 010008, Kazakhstan.}
\affiliation[e]{Mathematical and Physical Sciences Division, School of Arts and Sciences, Ahmedabad University, Ahmedabad 380009, India.}
\emailAdd{sbasak@iisertvm.ac.in}
\emailAdd{sukannya@physics.iitm.ac.in}
\emailAdd{mayukh@ctp-jamia.res.in}
\emailAdd{nurjaman@ctp-jamia.res.in}
\emailAdd{raghavan@ahduni.edu.in}
\emailAdd {sami$_{-}$ccsp@sgtuniversity.org; samijamia@gmail.com}

\abstract{In this paper, we consider a scenario  of spontaneous baryogenesis in a framework of warm quintessential inflation where the residual inflaton field, left out after warm  inflation, plays the role of quintessence field at late times and  is coupled to a non-conserved baryonic current. Assuming a four fermion $(B-L)$ violating effective interaction, we have demonstrated that the required baryon  asymmetry can be produced successfully in this case.  We show that the post-inflationary evolution, with the underlying scalar field potential, $V(\phi)=V^4_0 \exp{(-\alpha \phi^n/\mpl^4 )  }, n>1$ well suited to warm inflation, exhibits scaling behaviour soon  after a brief kinetic regime. We show that the coupling of the scalar field to massive neutrino matter can give rise to exit from the  scaling regime to cosmic acceleration at late times as massive neutrinos turn non-relativistic. The  proposed model is shown to successfully describe the cosmic history from inflation to late time acceleration, with the evolution independent of initial conditions, along with the generation  of baryon asymmetry during the post-inflationary era. A brief analysis of relic gravity waves produced in the scenario is presented.  }

\begin{document}
\maketitle

\section{Introduction}\label{intro}
The inflationary scenario is one of the most compelling paradigms evoked to address the inherent shortcomings of the standard model of the Universe. The framework not only addresses the inconsistencies associated with the standard cosmological model but has also proven to be one of the most promising answers to address the generation of 
cosmological perturbations, which evolve later as large scale structures and gravitational waves. The paradigm is supported by observations of the cosmic microwave background (CMB), such as Planck, WMAP etc.~\cite{ade2016planck,aghanim2018planck,hinshaw2013nine}. In the standard set up of inflation, a massive scalar field slowly rolling in the flat direction of its potential, induces (quasi)exponential cosmic expansion~\cite{Kazanas:1980tx,Guth:1980zm,Starobinsky:1980te, linde,shinji-rev}. In this standard scenario,  inflaton is treated as a very weakly coupled field  and thus the temperature remains negligible throughout inflation, mandating an additional epoch of reheating at the end of inflation, where the inflaton starts to oscillate at the bottom of the potential and decays at the end of inflation. This standard inflationary scenario with zero temperature is therefore termed as `cold inflation' (CI).


The basic idea of the WI scenario is that the inflaton  field  has non-negligible couplings with the pre-existing matter fields such as radiation during inflation \cite{berera,branden,hall-moss,Reyimuaji:2020bkm}. Due to the presence of such couplings, the inflaton can dissipate during the  process of inflation until the end of inflation characterised by either the slow-roll parameter(s) exceeding unity, or the radiation energy density ($\rho_{R}$) becoming dominant over the vacuum energy density($\rho_{V}$). Due to the inherent construction of WI, the temperature during inflation can no longer be neglected. Thus along with the quantum fluctuations of the inflaton field, in the WI scenario, one needs to consider the existence and effect of the thermal fluctuations as well \cite{oliviera,graham}. If inflation ends by $\rho_{R}$ dominating over $\rho_{V}$, it could end the accelerated expansion naturally~\cite{cerezo,berera2}. In addition to resolving theoretical problems mentioned above, the WI model resurrects the inflationary potentials which are otherwise ruled out in a CI paradigm~\cite{panotopoulos,bartrum,bastero-gil,arya2}.

One of the persistent and interesting questions in cosmology is the exact mechanism of the generation of baryon asymmetry in the early universe that would result in a prediction matching the observed value of the baryon asymmetry at present, $\eta _F \equiv \frac{n_b-n_{\bar{b}}}{s}\sim 10^{-10}$.
The presence of couplings of the inflaton with other relativistic fields motivates us to look for particular interactions that can lead to spontaneous baryogenesis even during inflation and soon thereafter. Spontaneous baryogensis \cite{Cohen:1988kt,Dolgov:1997qr} involves the  generation of baryon asymmetry of the Universe via coupling of the baryon current to the time derivative of some scalar field. In our scenario, the inflaton filed plays the role of time varying field.
Hence, given a particular model of WI and well-motivated baryon number violating interactions present during inflation, successful production of the observed baryon asymmetry would lead to constraints and better understanding of a WI scenario.

Another mystery in contemporary cosmology is related to the nature of the dark energy that governs the acceleration of the Universe at the present epoch. One of the well-inspired framework of dark energy is given by  quintessence models, where a scalar field rolling down a plateau potential drives the acceleration today. This motivates one to look for models of quintessential inflation~\cite{Peebles:1998qn}, where the inflaton does not decay completely after inflation, but survives till the present epoch to account for late time acceleration, remaining  invisible for most of the post inflationary history of the Universe from the end of inflation till the present epoch. The residual scalar field, can also be put to use for generation of the baryon asymmetry in the early Universe during inflation and in the subsequent  radiation/kinetic dominated era assuming its interaction with a non-conserved baryonic current, $\hat{\rm a}$  {\it  la} spontaneous baryogenesis. In this paper, we shall study spontaneous baryogenesis in the paradigm of quintessential inflation which was first proposed in the context of cold inflation~\cite{Peebles:1998qn}. It was latter studied in context of warm inflation~\cite{Dimopoulos:2019gpz, Lima:2019yyv}. Here we shall restrict to the latter scenario using a generalized exponential potential. We would invoke a coupling of the scalar field to massive neutrino matter for realising the  exit from scaling regime to late-time acceleration.

It is worthwhile to contrast warm quintessential
framework with its counterpart in the context of cold inflation. The latter requires an alternative reheating mechanism, for instance,
reheating due to gravitational particle production~\cite{Ford:1986sy,Chun:2009yu}, instant preheating~\cite{Felder:1998vq, Campos:2002yk}, curvaton reheating~\cite{Feng:2002nb, BuenoSanchez:2007jxm} and Ricci reheating~\cite{Dimopoulos:2018wfg, Opferkuch:2019zbd, Bettoni:2021zhq}. The gravitational particle production, in a sense, is distinguished,  it dose not require  additional fields unlike the other mechanisms except Ricci reheating where a non minimal coupling is present. However, it is quite inefficient; indeed, in this case, $\rho_r/\rho_\phi\simeq 10^{-15}$ \cite{Sahni:2001qp} at the end of inflation. Consequently, the field evolves in the kinetic regime for a long time before the commencement of radiation domination.   During the kinetic regime, the energy density in the relic gravity waves enhances compared to field energy density, thereby creating a challenge for primordial nucleosynthesis at the commencement of radiative regime. As for the other mechanisms mentioned, one chooses the model parameters for setting a lower bound on the reheating temperature
consistent with the nucleosynthesis constraint. The latter imposes a restriction on the duration of the kinetic regime.
Remarkably, it is a generic  feature of the
WI scenario that $\rho_\phi\sim \rho_r$ when inflation ends. Naturally, the kinetic regime is brief in this case posing no threat to the nuclesynthesis  due to relic gravity waves. Thanks to the said  distinguish feature, the WI paradigm better suits the framework of quintessential inflation compared to the standard cold inflation.

The rest of the paper is organised as follows. In the  section \ref{wi1}, we present a brief review of the WI dynamics and the analysis of the inflationary parameters. In section \ref{sbar}, we  discuss spontaneous baryogenesis and the corresponding analysis in the model under consideration. Then in section~\ref{lt}, we have discuss  late time implications of our model in details. Finally, the conclusions are drawn in section~\ref{conc}.

\section{ Warm inflation}\label{wi1}
\subsection{The model and evolution equations}
In the WI scenario, there is a continuous dissipation of energy from the inflaton field to the radiation bath, characterised by the dissipation coefficient $\Upsilon$, which, in general, depends on the temperature $T$ of the radiation bath and the inflaton field $\phi$. Such a dissipation offers additional friction to the field $\phi$, slowing it down even further than that of a CI scenario with slow roll. Therefore, the energy scale at which the relevant cosmological scales exit the inflationary horizon correspond to a lower energy scale in WI as compared to a CI scenarios with the same inflation potential $V(\phi)$. This leads to a smaller value of the observable tensor-to-scalar ratio $r$ in WI, which can lead to the rejuvenation of many large field models of inflation (e.g. $V(\phi)\propto~\phi ^4$), that are refuted by CMB data in  CI.
The equations governing the dynamics of WI are:
\begin{eqnarray}
\label{phid}
&\ddot{\phi}+3(1+Q)H\dot{\phi}+V_{,\phi}=0,\\
&\dot{\rho}_R+4H\rho _R= 3HQ\dot{\phi}^2.
\end{eqnarray}
Here, $Q\equiv \frac{\Upsilon}{3H}$, $\rho _R$ is the radiation energy density, and $V_{,\phi}\equiv \frac{dV}{d\phi}$. The Hubble parameter $H$ is given by the Friedmann equation:
\begin{equation}
\label{Hub1}
H^2 =\frac{1}{3\mpl ^2}\bigg(V(\phi)+ \frac{\dot{\phi}^2}{2}+\rho _R \bigg),
\end{equation}
where $\mpl = 2.38\times 10^{18}$ GeV is the reduced Planck mass.
Typically, one assumes the slow-roll condition to be valid throughout inflation, for which Eqs.~(\ref{phid})-~(\ref{Hub1}) can be further simplified. However, in a generic case, particularly when the energy budget near the end of inflation is important, the full equations should be considered for the evolution of $\phi$ and $\rho_R$. 

In terms of the number of e-folds $N$ from some fiducial pivot scale ($dN=Hdt$), these Eqs. can be written as:
\begin{eqnarray}
\label{phiN}
&\phi ''+\bigg(3(1+Q)+\frac{H'}{H}\bigg)\phi '+\frac{V_{,\phi}}{H^2}=0,\\
\label{rhoN}
&\rho _R'+4\rho _R = 3Q\phi '^2H^2,
\end{eqnarray}
and 
\begin{equation}
H^2=\frac{V(\phi)+\rho _R}{3\mpl ^2 -\phi '^2/2},
\label{HubN}
\end{equation}
where primes denote derivatives with respect to the number of e-folds $N$.
Thus, the dynamics of WI has two inputs: the inflaton potential $V(\phi)$ and the dissipation coefficient $\Upsilon (T,\phi)$. The exact functional form of $\Upsilon$ depends on the microphysics of the interactions of $\phi$ with the fields in the radiation bath~\cite{Berera:2008ar,Bastero-Gil:2019gao,Bastero-Gil:2010dgy,Bastero-Gil:2012akf,Bastero-Gil:2016qru}. The functional forms of $\Upsilon$ that are allowed by different types of such interactions are $\Upsilon \propto T^3/\phi ^2$, $\Upsilon \propto T$, and $\Upsilon \propto M^2/T $ ($M$ is a mass scale in the model). Thus, a generic form can be defined as:
\begin{equation}
\Upsilon = CT^c\phi ^p M^{1-c-p}
\label{Upsgen}
\end{equation}
\\
\indent In this work, we focused on quintessential warm inflation, where the same inflaton field $\phi$ leads to dark energy in late time. Such a scenario can be motivated by only a few forms of the potential $V(\phi)$. Ref.~\cite{Peebles:1998qn} explored a scenario where $V(\phi)\sim \lambda (\phi ^4 +M^4)$ leads to chaotic inflation and $V(\phi) \sim \lambda/(\phi ^4 +M^4)$ leads to  late time quintessence and these two parts of the potential are connected by a kinetic energy dominated phase, where the inflaton $\phi$ runs quickly towards the scaling regime. This model has been studied in the WI scenario in Ref.~\cite{Dimopoulos:2019gpz}. Here we are considering a particular model first introduced in Ref~ \cite{Geng:2015fla}:
\begin{equation}
V(\phi)=V_0 \exp \bigg(-\alpha \bigg(\frac{\phi}{\mpl}\bigg)^n \bigg),
\label{pot}
\end{equation}
and also explored in Refs.~\cite{Geng:2017mic,Ahmad:2019jbm, AresteSalo:2020yxl,Rezazadeh:2015dia}.  It can also lead to inflation in the early Universe and acceleration at late times, and this motivates us to consider the potential in Eq.~(\ref{pot} )in this work. This potential has been studied in the context of WI in Ref.~\cite{Lima:2019yyv}, under slow roll assumptions. Since the dissipative coupling of radiation with $\phi$ is only important at very early times and should not disturb the late time consequences for the model in Eq.~(\ref{pot}), we expect the form of $\Upsilon$ to be such that $Q$ is negligible soon after the end of inflation. It can be shown~\cite{Lima:2019yyv} that such a requirement leads to the choice of $c>2$ in Eq.~\eqref{Upsgen}. This motivates us to choose a dissipation coefficient that is cubic in the temperature:
\begin{equation}
\Upsilon = C_{\phi}\frac{T^3}{\phi ^2}.
\label{ups}
\end{equation}

\begin{figure}[b]
\begin{center}
\includegraphics[width=0.75\textwidth]{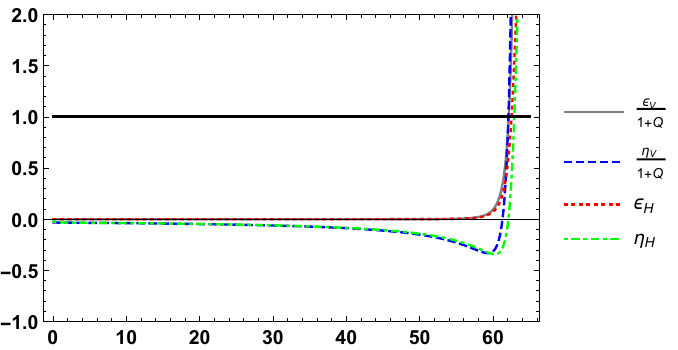}
\caption{{\small Evolution of the slow roll parameters with the number of e-folds. In this analysis, we track $\epsilon _H$ to determine the end of inflation.}}\label{srplots}
\end{center}
\end{figure}
Therefore, $Q=C_{\phi}\frac{T^3}{3H\phi ^2}$ and the full evolution equation for $Q$ is:
\begin{equation}
\frac{Q'}{Q}=-2\frac{\phi '}{\phi}-\frac{H'}{H}-3+\frac{9Q\phi '^2 H^2}{4\rho _R}.
\label{QN}
\end{equation}
Given the values of the model parameters, $\alpha $, $n$, $V_0$, and the initial conditions for $\phi$, $\phi '$, $Q$, $\rho _R$, we can solve Eqs.~(\ref{phiN}),~(\ref{rhoN}),~(\ref{HubN}),~(\ref{QN}) to arrive at the complete inflationary evolution. The evolution of the temperature $T$ during WI can be tracked using $T=\bigg(\frac{\rho _R}{(\pi ^2/30)g_*}\bigg)^{1/4}$, where $g_*$ is the number of relativistic degrees of freedom in the thermal bath. This analysis leads us to precise conditions at the onset of the post-inflationary epoch.
\subsection{Parameters for successful inflation}
The Hubble slow-roll parameters are defined as: 
\begin{eqnarray}
\epsilon _H&=& -\frac{H'}{H},\\
\eta _H &=& \epsilon _H - \frac{\epsilon '_H}{2\epsilon _H},
\label{SRH}
\end{eqnarray}
whereas, the potential slow-roll parameters are:
\begin{eqnarray}
\epsilon _V &=& \frac{\mpl ^2}{2}\bigg (\frac{V_{,\phi}}{V}\bigg )^2, \\
\eta _V &=& \mpl ^2 \frac{V_{,\phi \phi}}{V}.
\end{eqnarray}
However, in a WI setup, successful slow roll evolution requires $\epsilon _V/(1+Q), \eta _V/(1+Q)  \ll 1$.
We have considered: $V_0^{1/4} = 2.24\times 10^{15}$ GeV, $\alpha = 0.05$ and $n=3$. The values of different dynamical quantities at the pivot ($N=0$) are: $\phi _{\rm pivot}= 0.107 \mpl$, $\phi '_{\rm pivot} = 1.72\times 10^{-3} \mpl$, $Q_{\rm pivot}=3.20\times 10^{-6}$, and $\rho _{R, {\rm pivot}}^{1/4}=2.88\times 10^{12} $ GeV. 
Solving Eq.s~(\ref{phiN}),~(\ref{rhoN}),~(\ref{HubN}),~(\ref{QN}), we found that inflation ends when the kinetic energy comes to dominate the potential energy (slow roll violation), $\sim 62.8$ e-folds after the pivot scale leaves the horizon. This can be seen from the evolution of the slow-roll parameters plotted in Fig.~\ref{srplots}, where $\epsilon _H >1$ at $N=62.8$. It is evident from Fig.~\ref{srplots} that the inflaton follows exact slow-roll evolution for most of the inflationary epoch, apart from the last $\sim 2$ e-folds where $\epsilon _H$ and $\eta _H$ deviate considerably from $\frac{\epsilon _V}{1+Q}$ and $\frac{\eta _V}{1+Q}$ respectively. 
The inflationary observables are found to be consistent with CMB constraints from Planck 2018: the pivot scale amplitude of the scalar power spectrum: $\log (A_s\times 10^{10})=3.033$, scalar spectral index $n_s = 0.9616$, tensor-to-scalar ratio $r = 2.36\times 10^{-5}$.
\begin{figure}
\begin{center}
\includegraphics[width=0.9\textwidth]{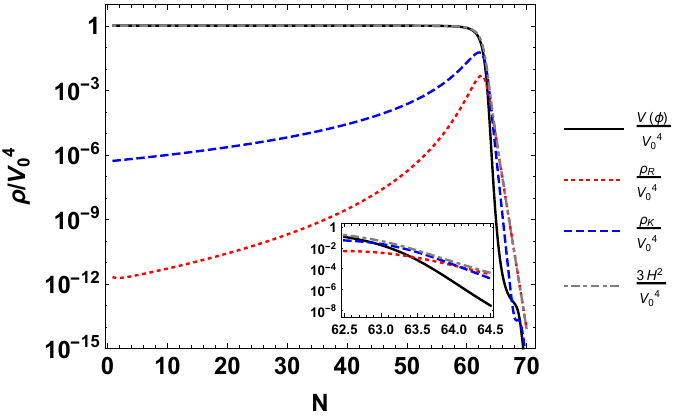}
\caption{{\small Evolution of the inflaton potential $V _{\phi}$ (solid black), radiation energy density $\rho_{\rm R}$ (dotted red) and kinetic energy density $\rho_{\rm K}$ (dashed blue) (all normalised with $V_0^4$) with the number of e-folds. The evolution of the $3H^2/V_0^4$ is also shown in dot-dashed grey curve. The inset shows crossover of energy densities near the end of inflation.}}\label{energyplots}
\end{center}
\end{figure}
The evolution of energy densities for the entire duration and close to the end of inflation are shown in Fig.~\ref{energyplots}. Evidently, after the end of inflation, the kinetic energy dominates the energy budget for $\sim 1.2$ e-folds, after which radiation domination begins.
\section{Spontaneous Baryogenesis}
\label{sbar}
\subsection{The model}
As we discussed in the introduction, the scalar field in this warm quintessential inflationary scenario survives after inflation. It goes through a very short kinetic epoch\footnote{This is contrary to the long kinetic epoch in quintessential CI with such a potential, which leads to interesting conclusions for baryogenesis and relic gravitational waves~\cite{Ahmad:2019jbm}. The kinetic epoch for the WI scenario here with the same potential is small because there is nonzero radiation energy density at the end of inflation here. In fact, we found $\rho_R\sim 0.1 V(\phi )$ at the end of inflation.} followed by a radiation dominated epoch. We imagine that the relevant interaction originates from a theory with a $U(1)$ symmetry, generated by some baryonic charge for the scalar field. Spontaneous breaking of a symmetry leads to  non-conservation of the baryonic current~\cite{Cohen:1988kt}.  In general, this kind of scenario involves derivative coupling of the scalar field with baryonic current in an effective field theory scenario with a cut-off characteristic scale for the $U(1)$ symmetry.  The effective Lagrangian can be written as~\cite{Dolgov:1997qr,DeSimone:2016ofp,  Ahmad:2019jbm, DeFelice:2002ir} 
\begin{eqnarray}
\mathcal{L}_{\rm eff}= \frac{\lambda '}{M}\partial _{\mu}\phi J^{\mu},
\label{Leff} \, ,
\end{eqnarray}
where $\lambda'$ is the coupling to baryonic charge, $M$ is the cut-off for the effective theory and $J^\mu$ is the non-conserved baryonic current. For a homogeneous FLRW Universe, Eq. (\ref{Leff}) takes the form, 
\begin{eqnarray}
\mathcal{L}_{\rm eff}= \frac{\lambda '}{M}\dot\phi \Delta n \label{Leff1}\, ,
\end{eqnarray}
where, $J^0=\Delta n =n_b-n_{\bar{b}}$ is the net baryon number density. It has been argued \cite{Cohen:1988kt} that the coefficient of  $\Delta n=n_b-n_{\bar{b}}$ in this case, shifts the energy of baryons and anti-baryon relative to each other, so $ \frac{\lambda'}{M}\dot \phi$ can be interpreted as an effective chemical potential for baryon numbers if $\dot \phi$ varies slowly. Then in thermal equilibrium, in the presence of $B$ violating interactions,
\begin{eqnarray}
\label{deln}
\Delta n= n_b -n_{\bar{b}}= \frac{1}{6} \bar{g} \frac{\lambda'}{M}\dot{\phi} T^2~,
\end{eqnarray}
where $\bar{g}$ is the internal degrees of freedom for baryons. 
 The third Sakharov condition for baryogenesis, namely, departure from thermal equilibrium, is relaxed in spontaneous baryogenesis because the time variation of $\phi$ provides a dynamical breaking of CPT (see Sec. 6.9 of \cite{Kolb:1990vq}, for more detail).

More recently it has been argued \cite{Arbuzova:2016spc,Dasgupta:2018eha} that the chemical potential interpretation of $\dot{\phi}$ is not correct. However, a net baryon asymmetry is generated~$(\propto \frac{\lambda'}{M}\dot \phi T^2)$ in such scenario with an $\mathcal{O}(1)$ proportionality constant that depends on the model and the interaction of the baryons. Therefore, we use the expression given in Eq.~(\ref{deln}) as an approximate expression for baryon asymmetry generated in thermal equilibrium in the presence of slowly varying $\dot{\phi}$.

 The baryon production freezes when the process of non-conservation of baryon current falls out of equilibrium compared to the Hubble expansion. If the freeze-out occurs at temperature $T_F$, we have the freeze-out condition ($\Gamma_B (T_F)\approx H (T_F)$). Beyond this time, 
interaction rate $\Gamma_B$( for baryon number violating process) can no longer cope with the Hubble expansion of the Universe ($\Gamma_B< H$). 

Now in thermal equilibrium, the entropy density is given by
\begin{eqnarray}
s= \frac{2 \pi^2}{45} g_{*,s} T^3 \, ,
\end{eqnarray}
where $g_{*,s}\simeq g_*$ is the entropy degrees of freedom at temperature $T$.
Thus the baryon to entropy ratio at decoupling/freeze-out is given by
\begin{eqnarray}
\eta _F \equiv \frac{\Delta n}{s}\vert _{T=T_F} = 0.38 \lambda ' \frac{\bar{g}}{g_*}\frac{\dot{\phi}(T_F)}{M T_F}\, .
\label{etaf}
\end{eqnarray}
 It is evident from Eq. (\ref{etaf}) that the freeze-out value of baryon  number depends on the time derivative of the scalar field at freeze-out temperature $T_F$, i.e. $\dot{\phi}(T_F)$, which depends on the model of inflation under consideration and $T_F$ depends on the exact $B$ violating process.
\begin{figure}[h]
\begin{center}
\includegraphics[width=14cm, height=6cm]{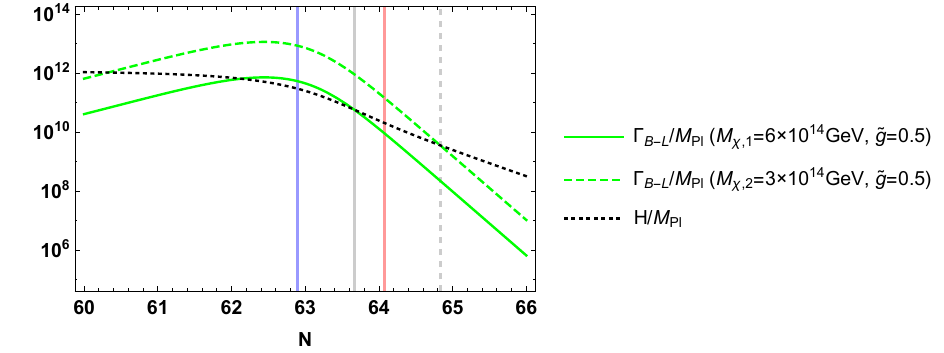}
\caption{{\small Freeze out of the baryon number violating process is shown for $M_{\chi}=6\times 10^{14}$ GeV (solid green) and $M_{\chi}=3\times 10^{14}$ (dashed green). The positions of freeze out in these two cases is shown in solid and dashed grey vertical lines respectively. The solid blue (extreme left) and solid red (third from left) vertical lines correspond to the onset of kinetic energy domination and radiation domination respectively.}}\label{gammaBLplot}
\end{center}
\end{figure}

The baryon number violating process is beyond the domain of the standard model of particle physics. However,  this can be realized in  the standard model  through non-renormalizable operators with a cut-off scale that violates the anomaly free combination $B-L$. To this effect, we consider the following $4-$fermi interaction
\begin{eqnarray}
\mathcal{L}_{B-L}=\frac{\tilde{g}}{M_{\chi}^2}\psi _1\psi _2\bar{\psi}_3\bar{\psi}_4 \label{fourfermiint}\, ,
\end{eqnarray}
where, $\psi_i$ are the fermions and $\tilde{g}$ is the coupling constant obtained  after integrating out the $B-L$ violating effects of particle of mass $M_{\chi}$. The rate of the process for $T< M_\chi$ is given by \cite{Kolb:1990vq}  
\begin{equation}
\Gamma _{B-L}(T)=\frac{\tilde{g}^2}{M_{\chi}^4}T^5,
\label{gambl}
\end{equation}
 This process freezes out at the temperature $T_F$ such that $\Gamma _{B-L}(T_F)=H(T_F)$.
\subsection{Analysis} 
For our analysis, we obtain the freeze-out condition for two values of $M_\chi$ namely,  $M_{\chi , 1}=6\times 10^{14}$ GeV and $M_{\chi , 2}=3\times 10^{14}$ GeV for $\tilde{g}=0.5$. From Eq.~\ref{gambl}, we found that for $M_{\chi , 1} $ the freeze-out happens during the kinetic energy dominated epoch, 1.14 e-folds after the end of inflation; whereas for $M_{\chi , 2}$ freeze-out takes place during the radiation dominated epoch, 2.31 e-folds after the end of inflation (see Fig.~\ref{gammaBLplot}). The freeze-out temperature and Hubble parameter at freeze-out in these two cases are $T_{F,1}= 10^{14}$ GeV, $H(T_{F,1})=6\times 10^{10}$ GeV and $T_{F,2}=4\times 10^{13}$ GeV, $H(T_{F,2})=4\times 10^9$ GeV respectively. Using the values $\bar{g}=2$, $\lambda '=10^{-5}$ and $M=10V_0$, we find $\eta _{F,1} = 4\times 10^{-9}$ and $\eta _{F,2} = 4\times 10^{-10}$ respectively for the two values of $M_{\chi}$ quoted above. We note that $\eta _{F,2}$ is in the right ballpark of the observed baryon asymmetry, however, $\eta _{F,1}$ is also viable, assuming that there can be late time entropy production before nucleosynthesis, which can further reduce $\eta _{F,1}$ to bring it inside the observationally allowed window.

\begin{figure}[tbph]
\begin{center}
$%
\begin{array}{c@{\hspace{.1in}}cc}
\includegraphics[width=3.5 in, height=3.5 in]{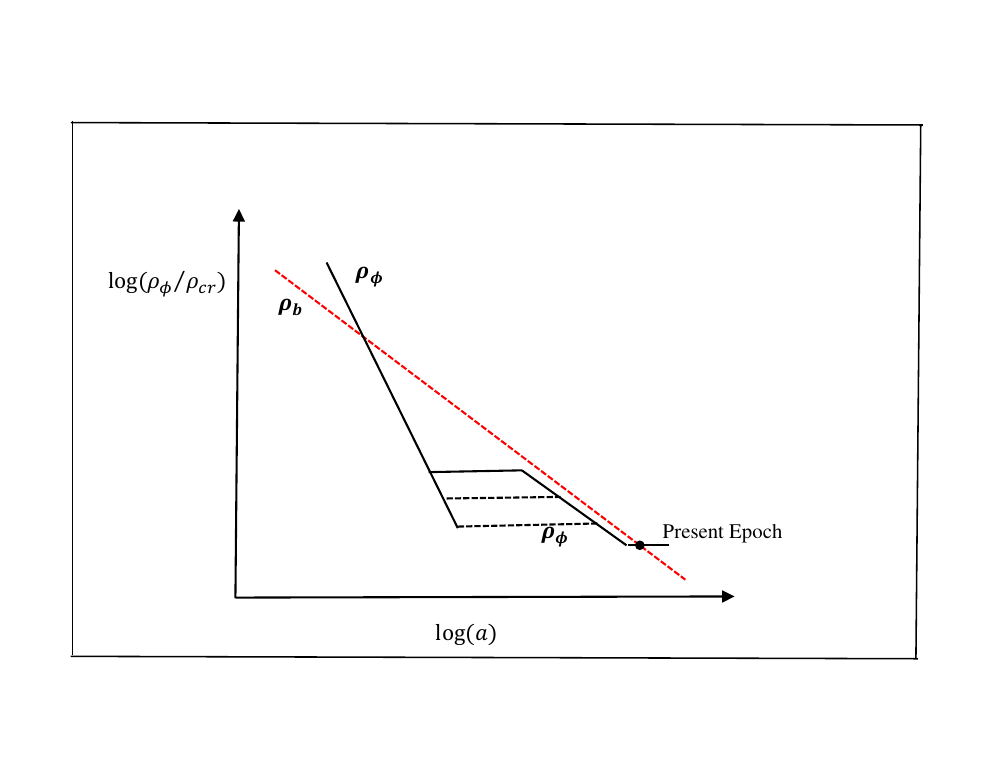} & %
\includegraphics[width=3.5 in, height=3.5 in]{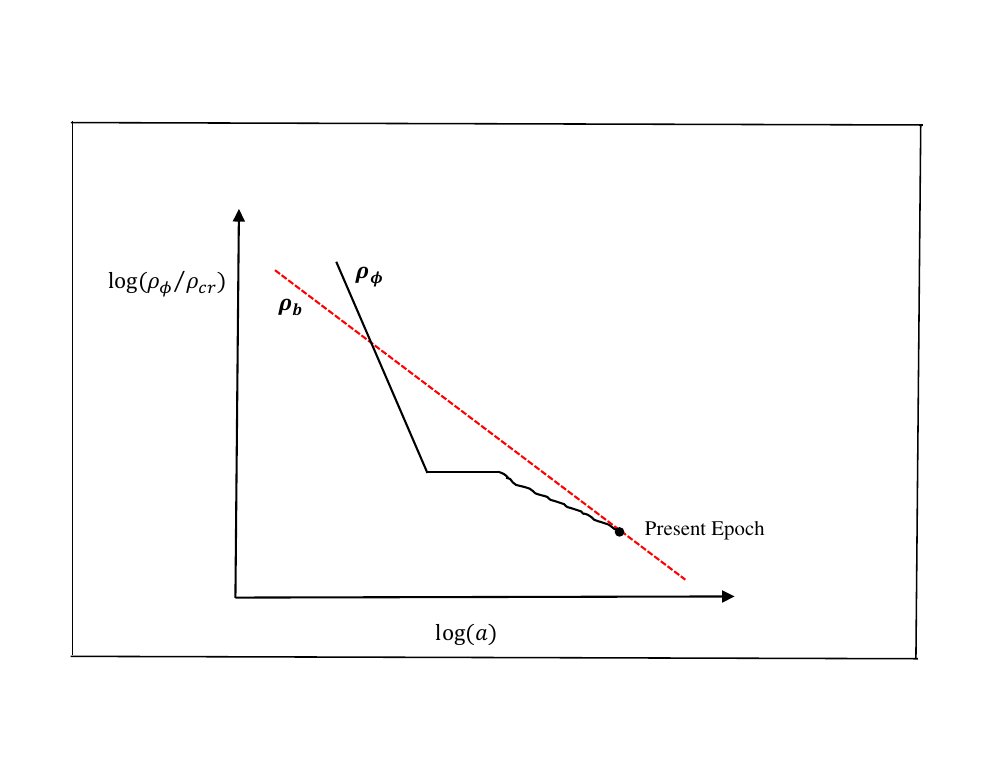} \\%
\mbox (a) & \mbox (b) &%
\end{array}%
$%
\end{center}

\begin{center}
$%
\begin{array}{c@{\hspace{.1in}}cc}
\includegraphics[width=3.5 in, height=3.5 in]{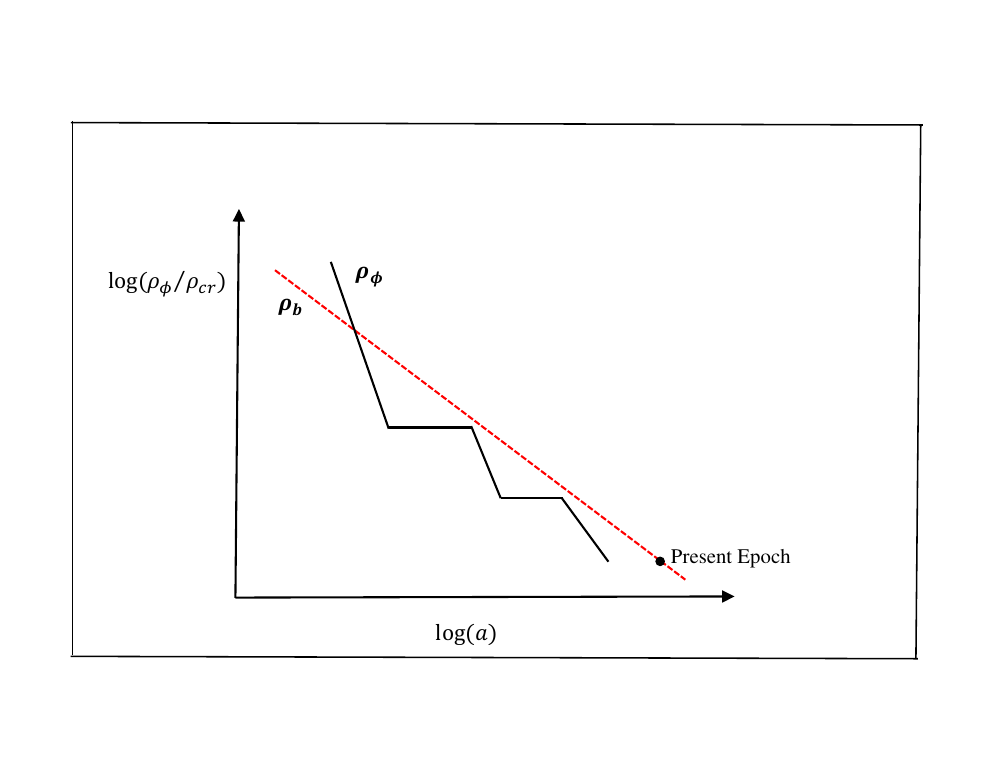} & %
\includegraphics[width=3.5 in, height=3.5 in]{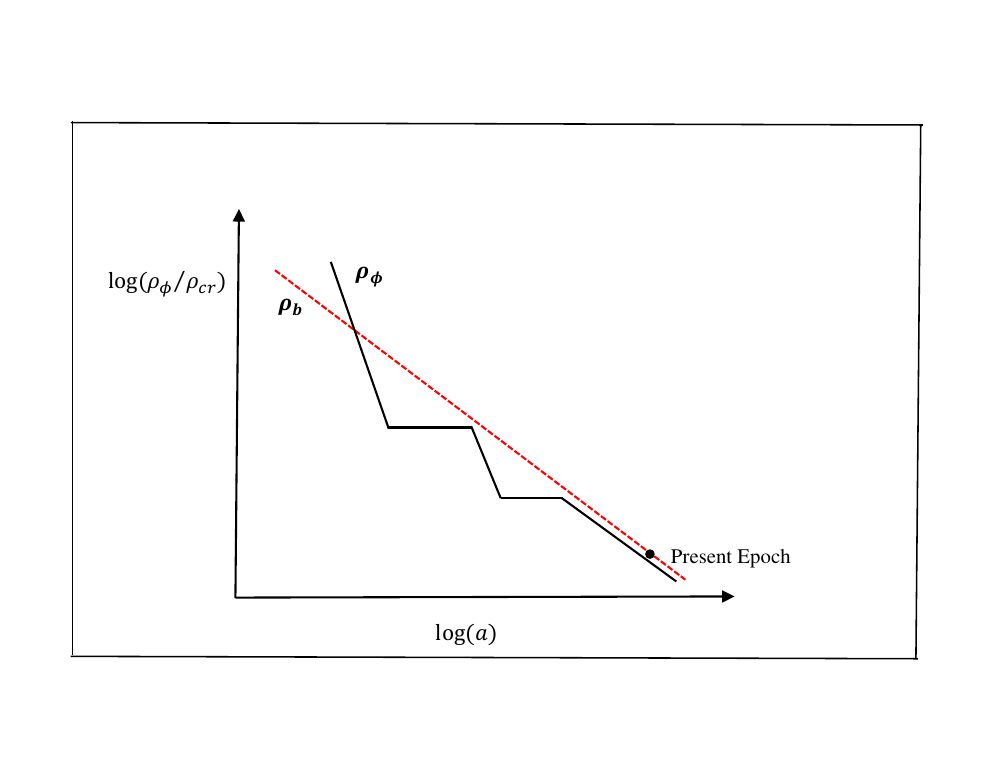} \\
\mbox (c) & \mbox (d)%
\end{array}%
$%
\end{center}

\caption{{\small The figures show the qualitative behaviour of the evolution of $\rho_\phi$ versus the scale factor on the log scale for a steep potential $V(\phi)$. The dotted line corresponds to the background (radiation/matter) energy density $\rho_b$. After the overshoot of $\rho_\phi$, the field freezes on its potential due to Hubble damping. After the recovery from freezing, field evolution crucially depends upon the nature of steepness. (a) In case of the exponential potential, the field catches up with the background and tracks it forever. Fig.~(b)  corresponds to the potential less steep than the exponential function.
Fig.~(c) exhibits the general feature of scalar field dynamics for a scalar field potential steeper than the exponential potential.
Fig.~(d) shows the evolution of $\rho_\phi$ for the generalized exponential potential in Eq.~(\ref{pot}) which dynamically mimics the exponential behaviour asymptotically.}}
\label{evolution}
\end{figure}

\begin{figure}
\begin{center}
\includegraphics[width=0.75\textwidth]{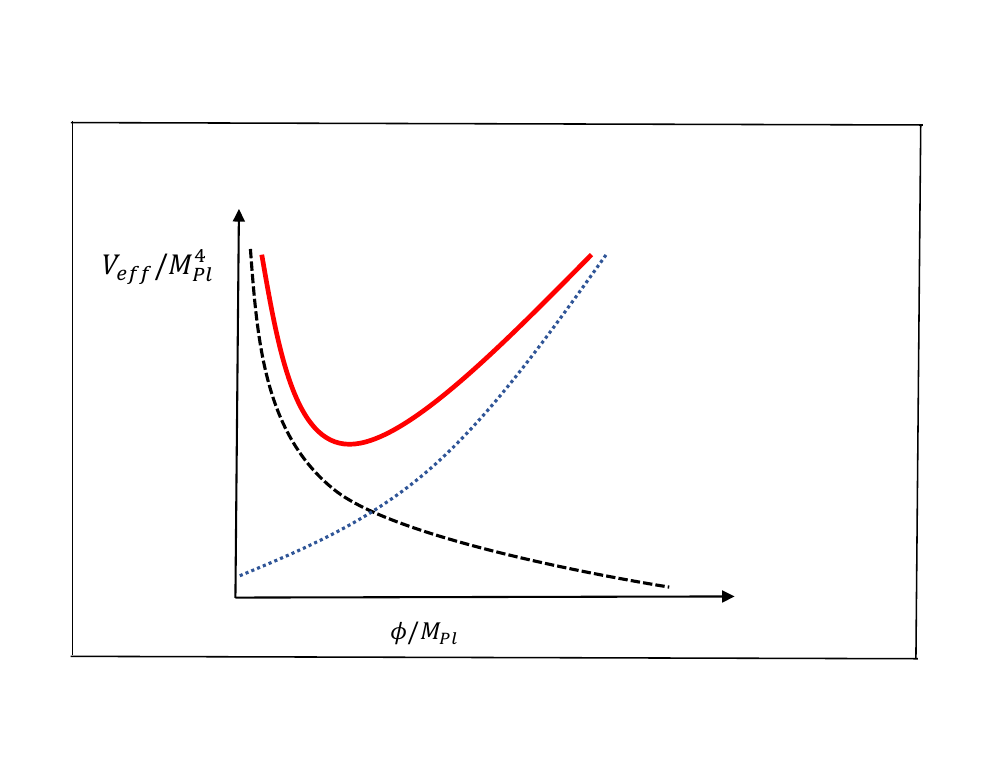}
\caption{{\small Qualitative picture of the effective potential (\ref{veffnu}). Dark dashed line is the original runaway potential (\ref{pot}); dotted line corresponds to $\hat{\rho_\nu} A(\phi)$. The effective potential has a minimum.}}
\label{effpot}
\end{center}
\end{figure}
\section{Late Time Cosmology}
\label{lt}
\subsection{Basics}
In the preceding sections, we have demonstrated that the WI scenario under consideration \cite{Lima:2019yyv,suratna}  successfully accounts for the observed baryon asymmetry of the Universe via spontaneous baryogenesis. In what follows we shall focus on the late time behaviour of the field $\phi$ and its
 role for dark energy. 
The choice of the potential in Eq. \eqref{pot} was made keeping in mind both the early and the late time evolution. The natural requirement
of the post-inflationary field dynamics is that the field does not interfere with the thermal history, $\hat{\rm a}$  {\it  la} nucleosynthesis, and the late time evolution is free from initial conditions.

The first condition requires a
steep field potential, whereas the second asks for a particular type of steepness. In this case, the field is dominant at early post inflationary stages for a very short period of kinetic domination, then the field energy density ($\rho_\phi$),  undershoots the background (radiation/matter) pushing the scalar field to freezing regime due to Hubble damping such that $\rho_\phi=const$ and $\phi$ freezes on its potential.
 Field evolution commences again as the background energy density becomes comparable to $\rho_\phi$. Hereafter, the deciding role is played
 by the nature of steepness of
the field potential. If $n=1$ in Eq.~\eqref{pot} (standard exponential potential), $\rho_\phi$ mimics the background (radiation/matter) dubbed scaling behaviour, see Fig.~\ref{evolution}(a). However, in case the field potential is less steep than the exponential one (inverse power-law behavior), $\rho_\phi$ gradually approaches the background and finally overtakes it (Fig.~\ref{evolution}(b)). On the other hand, if the potential is steeper than the standard exponential ($n>1$)~ {\color{magenta}\cite{Copeland:1997et, Copeland:2006wr, Haro:2019peq}}, the field energy density would evolve away from the
background, pushing the field into the freezing regime again,  after the recovery from which, the similar behavior repeats, see Fig.~\ref{evolution}(c).
This is a general feature
of scalar field evolution with potential steeper than the exponential in the FLRW space time.
In fact, the post inflationary behaviour of scalar field dynamics is controlled by a field construct, $\Gamma=V_{,\phi\phi} V/ V^2_{, \phi }$  {\color{ magenta} \cite{Steinhardt:1999nw}}, which is one for the exponential potential, giving rise to constant slope of the potential which is a  necessary criterion for scaling solution. Unfortunately, standard exponential potential is not suitable to inflation but the generalized exponential is. It is interesting to emphasize that the class of potentials in Eq.~\eqref{pot}  dynamically mimic the exponential behaviour in the asymptotic regime, namely $\Gamma \to 1$,  for large values of the field giving rise to a scaling solution which is an attractor of the dynamics~\cite{Skugoreva:2019blk}. In this case,  the field ultimately starts following (scaling) the background, see Fig.~\ref{evolution}(d).

Since the scaling solution is decelerating, we need a mechanism for late time exit from it to acceleration. 
As for the exit to late time acceleration, the simplest way out is provided by adding a cosmological constant to Eq.~\eqref{pot} or if we want the equation of state parameter different from minus one, we can add another exponential piece, $ \exp(-\lambda \phi/M_P), \lambda <\sqrt{2}$~\cite{Barreiro:1999zs} to the original potential such that its impact is felt only at late times. The latter, however, requires enormous amount of fine tuning. 
Another exit mechanism is provided by the
coupling of the field to cold matter~\cite{Lima:2019yyv, Amendola:1999er, Gumjudpai:2005ry} which induces a minimum in the  potential where the field could settle giving rise to a de Sitter solution, which, however, could spoil the matter regime if it happens at earlier times. Again it requires a huge amount of fine tuning to ensure that the field recovers from freezing only at late times and then evolves around the minimum.
This problem can be circumvented by invoking a coupling of massive neutrino matter to  the scalar field. By virtue of the tiny neutrino masses $\mathcal{O}(10^{-2})$ eV, they are relativistic at early times and their coupling to the field vanishes identically (coupling is proportional to the trace of the energy  momentum tensor of the massive neutrino matter). However, the coupling builds up dynamically at late times as the neutrinos turn non-relativistic, and this induces a minimum in the runaway potential \eqref{pot}, see Fig.~\ref{effpot}~\cite{Geng:2015fla}. 
 
\begin{figure}
\begin{center}
\includegraphics[width=0.65\textwidth]{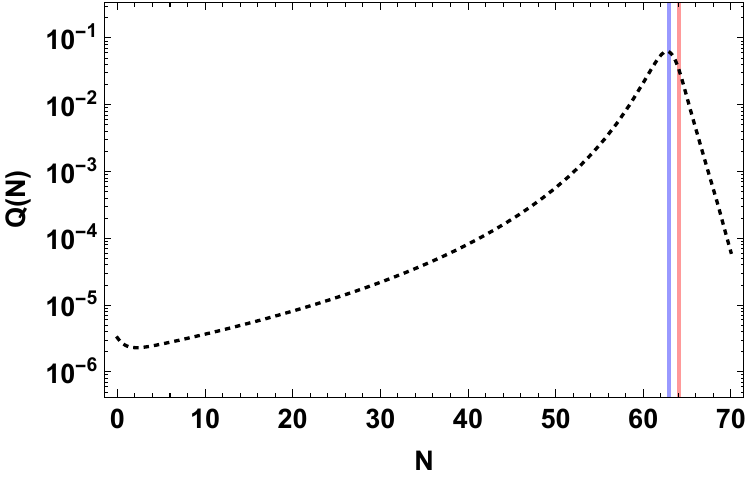}
\caption{{\small Evolution of $Q$ with the number of e-folds is plotted in black solid curve, where the solid blue (left) and solid red (right) vertical lines correspond to the onset of kinetic energy domination and radiation domination respectively.}}\label{Qplot}
\end{center}
\end{figure}

 Let us consider the action that includes
a coupling of $\phi$ to massive neutrino matter in the Einstein
  \cite{Hossain:2014xha,Geng:2015fla,Wetterich:2013jsa}\footnote{After inflation, the coupling of the radiation bath with the scalar field become ineffective, as $Q$ drops very sharply and has no role in later dynamics (see Fig.~\ref{Qplot}). So to incorporate the neutrino effect we have the action for nonminimal coupling (see Ref.~\cite{Geng:2015fla} and references therein).}, 
 \begin{eqnarray}
\mathcal{S}=\int d^4x \sqrt{-g}\left[ \frac{\mpl^2}{2}R -\frac{1}{2} \partial_\mu \phi \partial^\mu \phi -V(\phi)\right]+ \mathcal{S}_m +\mathcal{S}_r +\mathcal{S}_\nu \left( A^2(\phi) g_{\alpha\beta}, \right)\, ,
\label{nuaction}
\end{eqnarray}
where, $\mathcal{S}_m$ , $\mathcal{S}_r$ and  $\mathcal{S}_\nu $ are action for matter part, radiation part and neutrino part respectively.
The action \eqref{nuaction} implies direct coupling of field $\phi$ with  massive neutrino matter which is reflected in the field evolution equation as well as in the continuity equation for neutrino matter~\cite{Hossain:2014xha,Geng:2015fla,DeFelice:2010aj,samrev,Sami:2019hbq, Sami:2021ufn},\footnote{It may be noted that neutrino matter
is not coupled to $\phi$ in the Jordan frame and obey the standard continuity equation but gravity is modified, where as, in the Einstein frame, neutrino matter is coupled to the scalar field but gravity is standard and matter/radiation adhere to standard equation of continuity.}
\begin{eqnarray}
\label{frwconsnu}
 && \dot\rho_\nu+3H(\rho_\nu+p_\nu)=\frac{A_{,\phi}}{A}\dot\phi(-\rho_\nu+3p_\nu)\equiv  -\frac{A_{,\phi}}{A}\dot\phi \rho_\nu(3w_\nu-1)\\
 && \ddot\phi+3H\dot\phi+V_{,\phi}=-\frac{A_{,\phi}}{A}(-\rho_\nu+3p_\nu) \equiv \frac{A_{,\phi}}{A}\rho_\nu(3 w_\nu-1),
 \label{frwfdeqnu}
\end{eqnarray}
where $w_\nu$ denotes the equation of state parameter for massive neutrino matter. Eqs. (\ref{frwconsnu})  and (\ref{frwfdeqnu}) are obtained by varying the action (\ref{nuaction}) with respect to $g_{\mu\nu}$. During  most of the expansion history, neutrino matter behaves like radiation and their coupling to $\phi$ ($\propto T_\nu=-\rho_\nu+3p_\nu $) vanishes identically;  only at  late times when neutrinos turn non-relativistic, $w_\nu$ gradually changes from $1/3$ to zero and the coupling picks up non-zero values. We use the following ansatz \cite{Geng:2015fla, Hossain:2014xha} for the neutrino equation of state parameter,
 \begin{eqnarray}
 w_\nu (z)= \frac{1}{6}\Bigg\{1+ \tanh\left[\frac{\ln{1+z}-z_{\rm eq}}{z_{\rm dur}}\right]\Bigg\}\, \label{eq:wnu}
 \end{eqnarray}
 which interpolates between $1/3$ and $0$; here $z_{\rm eq}$ and $z_{\rm dur}$ are two parameters that determine respectively where and how fast the neutrino makes the transition from exhibiting radiation-like behaviour to cold matter-like behaviour.
 \begin{figure}[htbp]
\begin{center}
\includegraphics[width=0.8\textwidth]{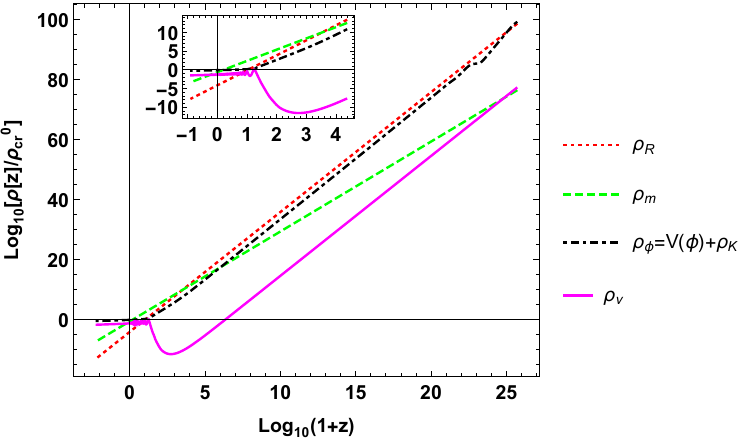}
\caption{{\small Evolution of the energy densities of matter (green dashed), radiation (red dotted), scalar field (black dot-dashed) and massive neutrinos (magenta), as a function of the redshift, in the case of the non-minimally coupled scenario, for $\alpha \gamma = 720$, $z_{\rm eq} = 2.54$ and $z_{\rm dur} = 2.93$ with the initial condition $\rho _{\nu}=10^{-22}\rho _{\rm cr}^0$ at the end of inflation. The parameters are chosen accordingly to match observation. Inset shows the oscillations in $\rho _{\nu}$ and scaling behaviour of $\rho _{\phi}$ near the present time.}}\label{rhoall_late}
\end{center}
\end{figure}

  As neutrinos turn non-relativistic around the present epoch, massive neutrino matter mimics cold  matter ($w_\nu$ gradually vanishes). Assuming  $w_\nu=0$, one might prefer to work with $\hat\rho_\nu=A^{-1} \rho_\nu$ which is conserved in the Einstein frame,
\begin{equation}
 \dot{\hat\rho}_\nu+3H\hat\rho_\nu=0 \, ,
 \label{wnu1}
\end{equation}
and the field evolution in the approximation under consideration acquires the following form,
\begin{equation}
 \ddot\phi+3H\dot\phi=-V_{,\phi}-A_{,\phi}\hat\rho_\nu ,
 \label{frwfeqd}
\end{equation}
which implies an effective potential for $\phi$,
\begin{eqnarray}
 V_{\rm eff}=V(\phi)+A(\phi)\hat{\rho}_\nu \,,
 \label{frwfeq}
 \end{eqnarray} 
 defined up to an additive constant.
Eq. (\ref{frwfeqd}) tells us that, in this case, the effect of coupling is incorporated solely in the effective potential which might be used for rough estimates. It should be noted that in case of coupling to standard cold matter, Eqs. (\ref{wnu1}) and (\ref{frwfeqd}) are exact in the matter dominated regime. In case of coupling to massive neutrino matter, these equations are approximate as $w_\nu$ is evolving, and for detailed investigations, one should use the set of coupled equations (\ref{frwconsnu}) and (\ref{frwfdeqnu}). 
Using a specific form of the conformal coupling, we have $A(\phi)=e^{\alpha \gamma \phi/\mpl}$ \cite{Geng:2015fla} so that
 \begin{equation}
 V_{\rm eff} \simeq V(\phi) + \hat{\rho}_{\nu} e^{\alpha \gamma \frac{\phi}{\mpl}}; ~\hat{\rho}_\nu=\rho_\nu e^{-\alpha \gamma \phi/\mpl}\,,
 \label{veffnu}
 \end{equation}
 where $\gamma$ is a constant 
 { and $\hat{\rho}_\nu= \rho_\nu$ for $\phi=0$}.

For the sake of illustration, we consider the case of $n=1$ in $V(\phi)$ in
Eq. \eqref{pot}.  Minimising $V_{\rm eff}$ with respect to $\phi$ we obtain analytically
\begin{equation}
\phi_{\rm min}=\log \bigg(\frac{V_0}{\gamma\hat{\rho}_\nu}\bigg) ^{1/\alpha(1+\gamma)}  \Rightarrow   V_{\rm eff}^{\rm min} \simeq V_0\left(\frac{\gamma \hat{\rho}_\nu}{V_0}\right)^{\frac{1}{(1+\gamma)}},
\label{phiminVeffmin}
\end{equation}
for moderately large value of $\gamma$ consistent with observations ($\gamma \sim 30)~$ \cite{Hossain:2014xha}. 
%
We used the fact that the effective potential is defined up to an additive constant and $1+\gamma\simeq \gamma$.We can estimate $V^{\rm min}_{\rm eff}$ at the present epoch to be identified with the dark energy density\footnote{ 
To obtain $\hat{\rho}_\nu$ in $V_{\rm eff}^{\rm min}$ and
$\phi_{\rm min}$ today we note that from Eq.~\eqref{phiminVeffmin}
\begin{eqnarray}
\frac{\hat{\rho}_\nu^0}{V_0}\simeq \frac{1}{\gamma}\left(\frac{V_{\rm eff}^{\rm min,0}}{V_0}\right)^\gamma\, 
\end{eqnarray}
Setting $V_{\rm eff}^{{\rm min},0}\simeq  H^2_0 \mpl^2=\beta \rho^0_\nu$,
where $\beta^{-1}\simeq \Omega^0_\nu$, and $V_0\approx M^4_{Pl}$ \cite{Hossain:2014xha}, we get
\begin{eqnarray}
\frac{\hat{\rho}_\nu^0}{\mpl^4}
\approx\left(\frac{\beta \rho^0_\nu}{M^4_{Pl}}\right)^\gamma\,.
\label{neutrinomatterdensities}
\end{eqnarray}
Then we get $\phi^0_{\rm min}/\mpl=\mathcal{O}(10)$.
Since $\rho_\nu$ grows exponentially from the inflationary
era to the present epoch, $\hat{\rho}_\nu^0$ is incredibly small. The relation of neutrino matter densities in Eq.~(\ref{neutrinomatterdensities}) is such that the evolution from the inflationary era to the present epoch is consistent with late time physics.}

The fixed point for the solution for scalar field domination corresponds to $\omega_\phi\simeq -1+3/{\alpha^2 \gamma^2}$ for large value of $\gamma$~\cite{Hossain:2014xha}. 

For $n=3$, numerical investigations indicate that
$\alpha\gamma$ should be large to get a (quasi) de Sitter solution.
%
%
$\alpha=0.05$ and $n=3$, the values we have chosen, are consistent with observational constraints related to the inflationary epoch~\cite{aghanim2018planck} in case of a generalized exponential potential. 
Our
numerical investigations of post-inflationary dynamics reveal that the consistency with late time cosmic acceleration implies that $ \gamma\simeq 15000$.
\subsection{ Late time exit from scaling regime to (quasi) de Sitter.}
The post-inflationary history of Universe is depicted in Fig.~\ref{rhoall_late} which shows the evolution of $\rho_\phi$, $\rho_R$, $\rho_m$ and $\rho_\nu$ versus the scale factor on the log scale, where $\rho _{\rm cr}^0$ is the critical energy density at present. The initial conditions for the late time evolution are taken from the onset of kinetic energy domination (KD), which are obtained from the inflationary evolution: $\phi_{\rm ini}^{\rm PI} = \phi \vert _{\rm KD} = 4.14 \mpl $ and $\dot{\phi}_{\rm ini}^{\rm PI} = \dot{\phi} \vert _{\rm KD} = 2.04\times 10^{-7} \mpl ^2$, where `PI' signifies that these are post-inflationary initial conditions. For modelling the neutrino equation of state $w_{\nu}$ using Eq.~(\ref{eq:wnu}), the values of $z_{\rm eq}=2.54$ and $z_{\rm dur}=2.93$ are chosen such that the values for the density parameters $\Omega _i$ match the observed values at present for all the components in the universe. At the end of inflation, $\rho_\phi$ and $\rho_r$ are within the same order of magnitude ($\frac{\rho_\phi}{\rho_R} \gtrsim 1$) in contrast to the cold inflation where $\frac{ \rho_{\phi}}{\rho_R}\vert _{\rm end}\gg1$. Consequently, in the present context, the overshoot of $\rho_\phi$ is small and the freezing regime is short. Soon after the recovery from freezing, the scalar field catches up with the background\footnote{With specific initial conditions, we have in the model after inflation, the ups and downs in $\rho_\phi$ shown in Fig.~\ref{evolution} are not visible here. } and follows it till it reaches the minimum of the effective potential where it oscillates and fast approaches the de Sitter state. The evolution of $\rho_\nu$ exhibits an interesting behaviour, especially at late times. Since neutrino matter is relativistic at early stages, $\rho_\nu$ tracks $\rho_R$ for most of the history. 
\begin{figure}[t]
\begin{center}
\includegraphics[width=.8\textwidth]{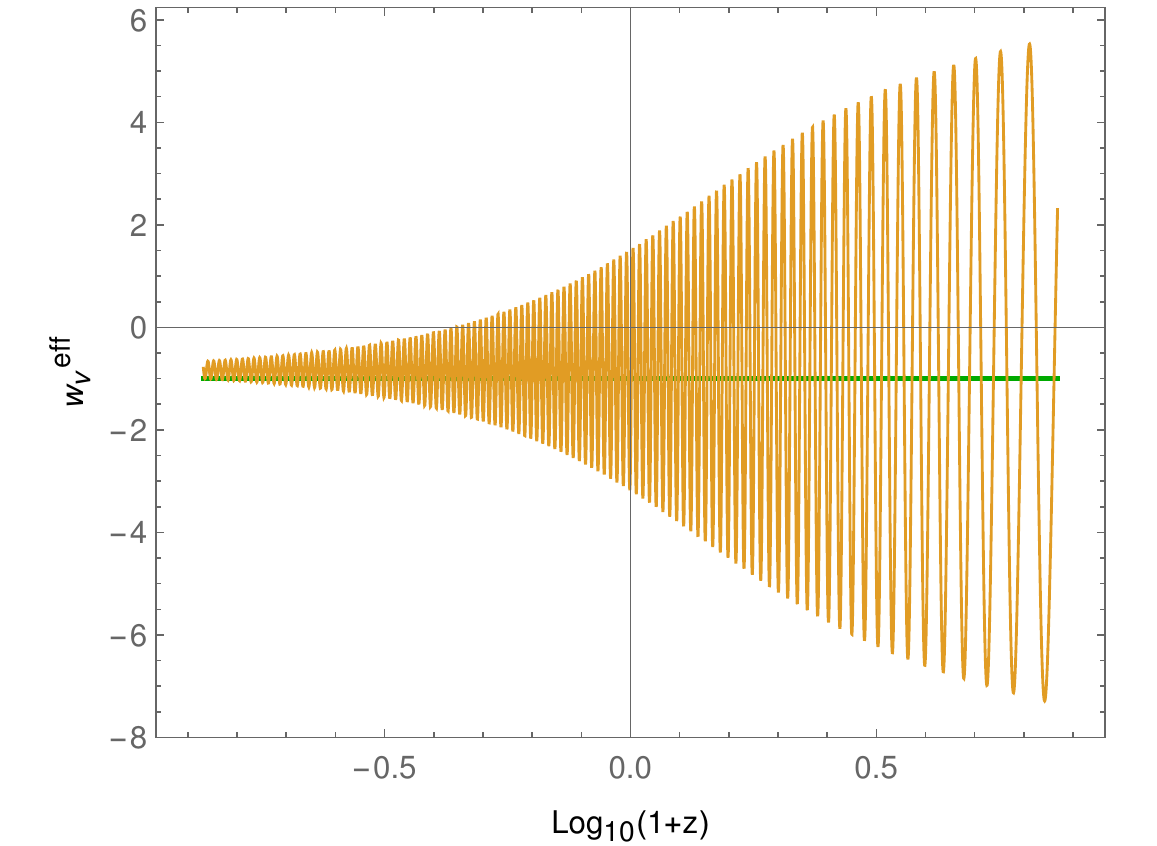}
\caption{\small Late time behaviour of ``$w_\nu ^{\rm eff }$" with $p_\nu=0$, we can see that it does oscillate about the average value close to $-1$, making neutrino density constant. The parameters value are same as in Fig.~\ref{rhoall_late}. This figure is based upon the simulation of the full dynamical system for $n>1$ without reference to effective potential.}
\label{wnuefffig}
\end{center}
\end{figure}
Near the present epoch, neutrino matter starts following the scalar field, namely, its energy density mimics the dark energy  like behavior ($\rho_\nu$ constant) which, at the onset, sounds counter intuitive. This behaviour can be explained recalling Eqs. (\ref{frwconsnu}) and (\ref{frwfdeqnu}). Taking into account the fact that  neutrino matter turns non-relativistic ($p_\nu\sim 0$) at late times,  Eq. (\ref{frwconsnu}) can be cast in the form,\\
\begin{eqnarray}
\dot \rho_\nu +3 H\rho_\nu \left(1- \alpha \gamma \frac{\dot\phi}{3 H \mpl}\right)=0\, .
\label{rhonueff1}
\end{eqnarray}
The last term in the bracket can be treated as an effective equation of state parameter for neutrino matter, 
\begin{equation}
\label{omnu0}
w_{\nu}^{\rm eff}\equiv  -\alpha \gamma \frac{\dot \phi}{3 H \mpl}.
\end{equation}
Numerical investigations  reveal that $\alpha \gamma$ should be large
for observational consistency forcing the equation of state parameter to oscillate around the average value, $w_{\nu}^{\rm eff}\simeq -1$ where it gradually settles, see Fig.~\ref{wnuefffig}. This  particular behaviour can be understood analytically for exponential potential ($n=1$). In  this case, we have the following fixed point of the underlying dynamical system \cite{Hossain:2014xha},
\begin{eqnarray}
\label{atr}
&&\omega_\phi=-\frac{\alpha^2\gamma(1+\gamma)}{3+\alpha^2\gamma(1+\gamma)};~~\omega_\nu=0\\
&& \dot\phi= \frac{3 H \mpl}{\alpha(1+\gamma)}
\end{eqnarray}
which is an attractor for $\gamma\geq 0$. The effective equation of state parameter for neutrino matter, $\omega^{eff}_\nu$ takes the following simple form at the fixed point given by (\ref{atr}),
\begin{equation}
 w_{\nu}^{\rm eff}=-\frac{\gamma}{1+\gamma}.
\label{omnu}
\end{equation}
It should be pointed out that the attractor solution is found in numerical investigations for $n=3$ using the full dynamical system without resorting to the effective potential picture.
Let us also note that $\gamma=0$ corresponds to the scaling solution in the matter era. In the absence of coupling, $\omega^{eff}_\nu=\omega_\nu=0$ as it should be.
From the effective potential perspective, in this case, the effective potential has no minimum and dynamics is dictated by the original steep potential responsible for the scaling solution.
For non-vanishing smaller values of $\gamma$, the minimum of the effective potential is shallow with small negative values of $w_\phi$ whereas the minimum is more pronounced for larger values of $\gamma$ giving rise to quasi de Sitter behavior \cite{Hossain:2014xha},
\begin{equation}
\label{avw}
\omega_\phi\simeq -1+\frac{3}{\alpha^2 \gamma^2}    
\end{equation}
which is
 followed by the neutrino matter as $w^{\rm eff}_\nu\simeq -1+1/\gamma \to -1$ for large values of $\gamma$. This kind of behaviour is  generic to models of coupled quintessence. In case of coupling to dark matter, $\omega_m$ exhibits the identical behaviour \cite{Gumjudpai:2005ry}. 
  One might naively think that in the late time quasi de Sitter phase, $\dot{\phi}\simeq 0$ and the effect of the coupling disappears on the RHS of Eq.(4.2). Then one would have $\rho_\nu \sim a^{-3}$. However, to obtain the quasi de Sitter phase at late time, one needs a large value of $\gamma$.  Now $\dot{\phi}\sim 1/\alpha \gamma$ ($\alpha>1$). 
Thus, the closer we are to the de Sitter phase, the smaller is $\dot{\phi}$.  However
$A_{,\phi}/A\sim\gamma$ and so we can not ignore the RHS of Eq.(4.2). As a result, neutrino matter density does not evolve as, $\rho_\nu\sim a^{-3}$.
Indeed, $w_\nu^{eff}=-\alpha \gamma \dot{\phi}/(3HM_{Pl}) \simeq -1$, see Eqs. (4.14) $\&$ (4.15), and Fig. 8. Thus neutrino matter does not evolve according to its own equation of state parameter, $\omega_\nu=0$ but rather follows the scalar field as the system approaches the quasi de Sitter phase at late time.

Let us reiterate that at early stages
 when neutrinos are relativistic $\omega_\nu=1/3$, coupling of the  field to neutrino matter is absent and the field follows the background (radiation/matter). However, 
 at late times as neutrinos gradually turn non-relativistic ($w_\nu\to 0$), the coupling builds up dynamically giving rise to a minimum in the potential such  that $\phi_{\rm min}(\hat{\rho}_\nu)$ 
 is time dependent  and so is $V_{\rm eff}^{\rm min}(\hat{\rho_\nu})$. 
 Because of the presence of the minimum, the
 field enters the oscillatory regime and oscillates with a negative average equation of state parameter given by Eq.~(\ref{avw}) for the given large value of $\gamma$.
 The field energy density, $\rho_\phi$ then starts red-shifting slower than the background (matter) energy density,  exits the scaling regime, takes over the background and becomes constant as the (quasi) de Sitter attractor is reached.
 Numerical investigations confirm similar behaviour for $n>1$, see  Fig.\ref{rhoall_late}. Clearly, the late time dynamics is dictated by the appearance of minimum in the effective potential  thanks to the late time coupling of neutrino matter to the scalar field.



\begin{figure}[t]
\begin{center}
\includegraphics[width=0.75\textwidth]{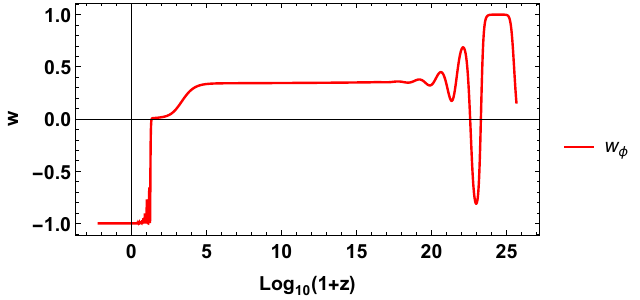}
\caption{{\small Evolution of the scalar-field equation-of-state parameters in the case of the non minimally coupled scalar field with neutrino  is shown  for $\alpha \gamma = 720$, $z_{\rm eq} = 2.54$ and $z_{\rm dur} = 2.93$ with the initial condition $\rho _{\nu}=10^{-22}\rho _{\rm cr}$ at the end of inflation.}}\label{wall_late}
\end{center}
\end{figure}

\section{Conclusions}
\label{conc}
In today's era of precision cosmology, theoretically motivated models of the primordial Universe (e.g. models of inflation ) are continuously put to test with the cosmological data. On the other hand, state-of-the-art observational surveys of CMB, large scale structures, gravitational waves and several stellar objects are conferring with exquisite theoretical models to answer some of the persistent questions in contemporary cosmology. In this work, we have constructed a viable warm inflationary scenario, which (i) explains the observed baryon asymmetry via the mechanism of spontaneous baryogenesis during the last few e-folds of inflation and for a short duration thereafter; and (ii) explains late time acceleration using the same scalar field. 

All through the primordial, post-inflationary and late universe evolution of our setup, one can identify specific points of high phenomenological relevance:(I) inflationary pivot point ($N=0$); (II) freeze-out of the B-L violating interaction ($T=T_F$); and (III) present epoch ($z=0$). The warm quintessential inflation scenario with late-time non-standard neutrino coupling considered here is subject to several constraints and insights implemented by the observational quantities at these points of evolution. 

(I) The model of quintessential inflation in the warm setup has the following set of parameters: $n$, $\alpha$ and $V_0$. CMB surveys, such as Planck, constrain the amplitude and scale dependence of scalar and tensor perturbations at the time when a pivot scale $k=0.002$ Mpc$^{-1}$ leaves the inflationary horizon. As explained in the Analysis of Section~\ref{wi1}, with $n=3$, $\alpha = 0.05 $, we could achieve the scalar spectral index $n_s=0.962$ and tensor-to-scalar ratio $r=2.36\times 10^{-5}$, which are within the confidence limit given by Planck 2018: $n_s=0.9649\pm 0.0042$ ($68\%$ confidence limit for the data combination Planck TT,TE,EE
+lowE+lensing) and $r < 0.056 $ ($95\%$ confidence limit for the data combination Planck TT,TE,EE
+lowE+lensing+BK15). $V_0=2.24\times 10^{15}$ GeV is fixed with $\log (A_s\times 10^{10})=3.03$, which is within $1\sigma$ confidence limit of the observed amplitude $\log (A_s\times 10^{10})=3.044\pm 0.014$ ($68\%$ confidence limit for the data combination Planck TT,TE,EE
+lowE+lensing).\\
\begin{figure}
\begin{center}
\includegraphics[width=0.75\textwidth]{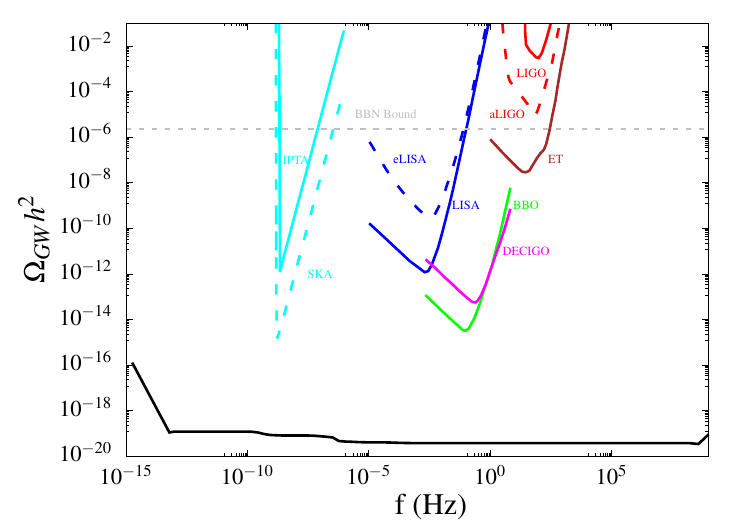}
\caption{{\small Energy density of the gravitational wave spectrum. Sensitivities of present observations and proposed sensitivities of the future detectors are also shown.}}\label{GW}
\end{center}
\end{figure}

(II) In the particular scenario under consideration, the WI paradigm comes to an end when the kinetic energy $\rho _K$ of the inflaton dominates over the potential energy, i.e., the energy hierarchy is $\rho_K>V(\phi)>\rho_R$ and $\epsilon_H=1$. Interestingly, the kinetic energy dominated epoch lasts for a very short period of time ($\sim 1.2$ e-folds) after which radiation domination begins. Thus, the post-inflationary evolution here is very close to the standard evolution. But, the scalar field takes part in the B-L violating interactions and spontaneous baryogenesis takes place during the end of inflation and freezes out in the post-inflation regime, as is evident from the evolution of $\Gamma _{B-L}$ and $H$ in Fig.~\ref{gammaBLplot}. The parameters that predict the theoretical value for the baryon to entropy ratio $\eta _F$ are $\lambda '$ and $M$ associated with the coupling of the baryon current to $\dot{\phi}$ in Eq.~\eqref{Leff}, and on $\tilde{g}$, $M_{\chi}$ associated with the $B-L$ violating effective interaction as in Eqs.~\eqref{fourfermiint} and~\eqref{gambl}. The value $\tilde{g}=0.5$ is taken for a standard coupling and $M_{\chi}$ is the characteristic scale of the interaction. The two values of $M_{\chi}$ are chosen such that $M_{\chi}>T_{\rm end}$. Therefore, the $B-L$ violating effective interaction in Eq.~\eqref{fourfermiint} is valid during inflation. $M=10V_0$ is chosen judiciously so that the derivative coupling associated with a spontaneously broken symmetry is valid at the inflationary scale $V_0$. The extremely short kinetic domination epoch along with very small value of the tensor-to-scalar ratio during inflation amount to very feeble relic gravitational waves (GW) $\Omega _{\rm GW}h^2\sim \mathcal{O}(10^{-19})$, which grows for a very small window in frequencies around $f\sim \mathcal{O}(10^8)$ Hz, which can be seen in Fig.~\ref{GW} (see Appendix I for details).

(III) Since the scalar field in this case is quintessential in nature with a nonstandard steep potential ($n=3$), the discussion in Sec.~\ref{lt} clarifies that the energy density of the field, $\rho _{\phi}$ tracks the background energy density from the end of kinetic domination until when the coupling with massive neutrinos given in Eqs.~\eqref{frwconsnu} and~\eqref{frwfdeqnu} becomes relevant at late times (see Fig.~\ref{rhoall_late}). As the neutrinos gradually become non-relativistic over a duration of e-folds $z_{\rm dur}$, the coupling to the scalar field velocity $\dot{\phi}$ makes $\rho _{\nu}$ to grow, and eventually settle at a constant value. This behavior of $\rho _{\nu}$ at late times  tracks  $\rho _{\phi}$, which  dominates the late-time energy density with $\rho _{\phi}\simeq$ constant. The relevant coupling between the scalar field and neutrino sector here is controlled by the parameter $\gamma$ explicitly, and implicitly by $z_{\rm eq}$ and $z_{\rm dur}$. The value $\alpha\gamma =720$ considered here is large, which is explained as a requirement in Sec.~\ref{lt}. The initial conditions for $\phi$ evolutions at late times is obtained from the early Universe evolution, since the initial time for studying the late time cosmology is taken to be at the onset of kinetic domination. The initial neutrino energy density $\rho _{\nu}$ acts as a normalising factor to achieve the correct current densities $\rho _{\nu}^0$ and $\rho _{\phi}^0$.

In this work, we have achieved the goal to successfully explain observational signatures of inflation, baryon asymmetry and dark energy density with a model of warm quintessential inflation. Previous well-motivated attempts in baryogenesis from warm inflation were made in Ref.~\cite{branden} with a large-field potential $V(\phi)=\frac{1}{2}m^2\phi ^2$ and in Ref.~\cite{Bastero-Gil:2011clw} via a two-stage mechanism. However, these works did not incorporate quintessence dark energy. On the other hand, quintessential inflation has been studied in the WI paradigm in references~\cite{Lima:2019yyv,Dimopoulos:2019gpz} for different potentials. Our work attempts to combine these two approaches and successfully leads to a model which is consistent with the spectrum of primordial density perturbations, baryon asymmetry and present day acceleration as determined from observations.


\section*{Acknowledgement}
S.~Bhattacharya is supported by the National Postdoctoral Fellowship of the Science and Engineering Research Board (SERB), Department of Science and Technology (DST), Government of India (GOI).
Work of MRG is supported by Department of Science and Technology (DST), Government of India under the Grant Agreement number IF18-PH-228 (INSPIRE Faculty Award) and partially by Science and Engineering Research Board (SERB), DST, Government of India under the Grant Agreement number CRG/2020/004347 (Core Research Grant). MS is partially supported by the Ministry of Education and Science of the Republic of Kazakhstan, Grant
No. 0118RK00935 and by NASI-Senior Scientist Platinum Jubilee Fellowship (2021). The authors sincerely thank Prof. Mar Bastero-Gil and Prof. Rudnei Ramos for useful discussions regarding this work. MRG wants to thank Amogh K. Rai for help during the draft formatting. NJ is thankful to Richa Arya, Wali Hossain and Arvind Mishra for fruitful discussions.
We would also like to thank the referee for her/his comments which have contributed to enhancing the arguments made in this article.

\section*{Appendix I: Relic Gravitational Waves}
\label{appendI}
Gravitational Waves (GWs) originate during inflation from the metric perturbations of the perturbed spatially flat FLRW universe  $ds^2= -dt^2 +a(t)^2\left( \delta_{ij} +h_{ij}\right) dx^i dx^j$, where the tensor perturbations $h_{ij}$ are transverse ($ \partial_i h_{ij}=0 $) and trace-less condition ($h_{ii}=0$). The GW spectrum is defined as \\
\begin{eqnarray}
\Omega_ {GW}(k, \tau)\equiv \frac{1}{\rho_{\rm cr} (\tau)}\frac{d \rho_{\rm GW}}{d\ln k}\, , \label{omegagw1}
\end{eqnarray}
where $\rho_{\rm GW}$ is the gravitational wave energy density and $\tau $ is the conformal time. Eq. \eqref{omegagw1} can be written  for today as \cite{Ahmad:2019jbm, Figueroa:2018twl}
\begin{eqnarray}
\Omega_{\rm GW,0}=\frac{1}{12}\left( \frac{k^2}{H_0^2 a_0^2}\right) \Delta_T^2 (k) T^2 (k,\tau)\, , 
\end{eqnarray}
where $\Delta_T^2 (k)= \left. \frac{2}{\pi^2} \frac{H_{\rm inf}^2}{\mpl^2}\right |_{k=aH}$ is the primordial tensor power spectrum at horizon crossing of the mode $k$ during inflation and $T^2 (k,\eta)$ is the transfer function. The transfer function describes the gradual evolution of GWs from the point of horizon entry of a mode $k$ in the post-inflationary universe until the time $\tau$ of observation. The transfer function today ($\tau = \tau _0$) is given as  \cite{Kuroyanagi:2008ye,Boyle:2005se,Watanabe:2006qe}
\begin{eqnarray}
T^2(k)\approx \frac{1}{2} \frac{a_{\rm hc}^2}{a_0^2}\, , \label{TansferK}
\end{eqnarray}
where the subscript  $ ``0"$ denotes present time and $``\rm hc"$  represents horizon crossing after inflation. The GW spectrum today for modes entering at different post-inflationary epochs of kinetic domination (KD), radiation domination (RD) and matter domination (MD) can be written as \cite{Ahmad:2019jbm, Figueroa:2018twl}:
\begin{eqnarray}
\label{omgwrd}
\Omega_{\rm GW}^{\rm RD, 0}(k)&=& \frac{1}{6 \pi^2} \Omega_{\rm R}^0 \frac{H_{\rm inf} ^2}{\mpl^2}\left(\frac{g_{*}(a_k)}{g_{*0}}\right) \left(\frac{g_{*,s}(a_k)}{g_{*,s0}}\right)^{-4/3}~ ~ {\rm for}(k_{\rm RM}<k<k_{\rm KR})\\
\label{omgwkd}
\Omega_{\rm GW}^{\rm KD, 0}(k)&=& \tilde{\Gamma}(\alpha _w) \Omega_{\rm GW}^{\rm RD, 0} \frac{a_kH_k}{a_{\rm KR}H_{\rm KR}}~ ~ ~ ~ ~ ~ ~ ~ {\rm for}(k > k_{\rm KR})\\
\label{omgwmd}
\Omega_{\rm GW}^{\rm MD, 0}(k)&=& \frac{1}{6 \pi^2} \Omega_{\rm m}^0 \frac{H_{\rm inf} ^2}{\mpl^2}~ \frac{H_0^2}{a_k^2H_k^2} ~ ~ ~ ~ ~ ~ ~ ~{\rm for}(k < k_{\rm RM}).
\end{eqnarray}

Here, the subscript `$k$' signifies quantities computed at the horizon entry of the mode $k=a_kH_k$, the subscript `KR' signifies quantities evaluated at the transition from KD to RD, `RM' implies the mode at matter-radiation equality. $\Omega _R^0$ and $\Omega _m^0$ are the dimensionless energy densities for radiation and matter at present. The quantity $\tilde{\Gamma}(\alpha _w)\equiv \bigg( \frac{\Gamma (\alpha _w +1/2)}{\Gamma (3/2)}\bigg)^2\frac{1}{2^{2-\alpha _w}\alpha _w ^{2\alpha _w}}\sim \mathcal{O}(1)$ is introduced for gradual transition from KD to RD epoch~\cite{Figueroa:2018twl}, where $\alpha _w\equiv\frac{2}{1+3w}\vert_{w=1}=1/2$.  The Hubble parameter after inflation is always computed here using 
\begin{equation}
3H^2\mpl ^2 = \rho _{\phi} + \rho _R + \rho _m + \rho _{\nu},
\end{equation} 
where each of the quantities are discussed in Sec.s~\ref{wi1},~\ref{sbar} and~\ref{lt}, and evaluated throughout cosmological evolution.

It is evident from the Eqs.~\eqref{omgwrd},~\eqref{omgwkd} and~\eqref{omgwmd} that for a nearly constant $H_{\rm inf}$, as is the case in this work, the GW energy spectrum for RD is nearly constant for the range  $k_{\rm RM} - k_{\rm KR}$, whereas the GW spectrum has a blue-tilt for the modes entering in the KD epoch and a red-tilt for the modes entering in the MD epoch(A detailed study of the GW production in case of warm inflation can be found in the references~\cite{mrgsarma, mbggw}). The full spectrum is shown in Fig.~\ref{GW}, where the frequencies are calculated as $f=2\pi k$ and the sensitivity curves for present and upcoming observations are also shown\footnote{Plotted using \href{gwplotter}{http://gwplotter.com/} based on Ref.~\cite{Moore:2014lga}.}. We find that the full GW spectrum for the scenario considered in this work stays much smaller than the present and future observational sensitivity curves owing to the small value of the tensor-to-scalar ratio $r\sim \mathcal{O}(10^{-5})$ at the pivot scale, since $\Delta _T^2\equiv r\Delta _S^2 = \frac{2}{\pi ^2}\frac{H_{\rm inf}^2}{\mpl^2}$, where $\Delta _S^2$ is the inflationary scalar power spectrum.


\begin{thebibliography}{99}
\bibitem{ade2016planck}
P. A. R. Ade {\emph et al}, Astron. \& Astrophys. {\bf 594} (2016) A13
[arXiv:1502.01589 [astro-ph.CO]].

\bibitem{aghanim2018planck}
N.~Aghanim \textit{et al.} [Planck],
Astron. Astrophys. \textbf{641}, A6 (2020)
[erratum: A                  stron. Astrophys. \textbf{652}, C4 (2021)]
[arXiv:1807.06209 [astro-ph.CO]].

\bibitem{hinshaw2013nine}
G. Hinshaw {\emph et al}., Astrophys. J. Suppl. {\bf 208} (2013) 20
 	[arXiv:1212.5225 [astro-ph.CO]].

\bibitem{Kazanas:1980tx}
D.~Kazanas,
Astrophys. J. Lett. \textbf{241}, L59-L63 (1980).
\bibitem{Starobinsky:1980te}
A.~A.~Starobinsky,
Phys. Lett. B \textbf{91}, 99-102 (1980).
\bibitem{Guth:1980zm}
A.~H.~Guth,
Phys. Rev. D \textbf{23}, 347-356 (1981).
\bibitem{linde}
A. D. Linde, Phys. Lett. {\bf 108B} (1982) 389.

\bibitem{shinji-rev}
S.~Tsujikawa,
[arXiv:hep-ph/0304257 [hep-ph]].

\bibitem{berera}
A.~Berera,
Phys. Rev. Lett. \textbf{75}, 3218-3221 (1995)
[arXiv:astro-ph/9509049 [astro-ph]].
\bibitem{branden}
R.~H.~Brandenberger and M.~Yamaguchi,
Phys. Rev. D \textbf{68}, 023505 (2003)
[arXiv:hep-ph/0301270 [hep-ph]].

\bibitem{hall-moss}
L.~M.~H.~Hall, I.~G.~Moss and A.~Berera,
Phys. Rev. D \textbf{69}, 083525 (2004)
[arXiv:astro-ph/0305015 [astro-ph]].
\bibitem{Reyimuaji:2020bkm}
Y.~Reyimuaji and X.~Zhang,
JCAP \textbf{04}, 077 (2021)
[arXiv:2012.07329 [astro-ph.CO]].
\bibitem{oliviera}
H.~P.~De Oliveira and S.~E.~Joras,
Phys. Rev. D \textbf{64}, 063513 (2001)
[arXiv:gr-qc/0103089 [gr-qc]].

\bibitem{graham}
C. Graham and I. G. Moss, (2009) JCAP, {\bf 2009}(07) (2009) 013
[arXiv:0905.3500 [astro-ph.CO]]
\bibitem{cerezo}
R. Cerezo and J. G. Rosa, JHEP {\bf 2013}(1) (2013) 24
[arXiv:1210.7975 [hep-ph]].


\bibitem{berera2}
A.~Berera,
PoS \textbf{AHEP2003}, 069 (2003)
[arXiv:hep-ph/0401139 [hep-ph]].

\bibitem{panotopoulos}
G. Panotopoulos and N. Videla,  Eur. Phys. J. {\bf C} {\bf 75}(11) (2015) 525
[arXiv:1510.06981 [gr-qc]].

\bibitem{bartrum}
S. Bartrum, M. Bastero-Gil, A. Berera, R. Cerezo,
R. O. Ramos and J. G. Rosa, Phys. Lett. {\bf B} {\bf 732} (2014) 116 
[arXiv:1307.5868 [hep-ph]].

\bibitem{bastero-gil}
M.~Bastero-Gil, S.~Bhattacharya, K.~Dutta and M.~R.~Gangopadhyay,
JCAP \textbf{02}, 054 (2018)
[arXiv:1710.10008 [astro-ph.CO]].
\bibitem{arya2}
R.~Arya, A.~Dasgupta, G.~Goswami, J.~Prasad and R.~Rangarajan,
JCAP \textbf{02}, 043 (2018)
[arXiv:1710.11109 [astro-ph.CO]].

\bibitem{Sakharov:1967dj}
A.~D.~Sakharov,
Pisma Zh. Eksp. Teor. Fiz. \textbf{5} (1967), 32-35.
\bibitem{Cohen:1988kt}
A.~G.~Cohen and D.~B.~Kaplan,
Nucl. Phys. B \textbf{308}, 913-928 (1988).
\bibitem{Dolgov:1997qr}
A.~D.~Dolgov,
[arXiv:hep-ph/9707419 [hep-ph]].
\bibitem{Peebles:1998qn}
P.~J.~E.~Peebles and A.~Vilenkin,
Phys. Rev. D \textbf{59}, 063505 (1999)
[arXiv:astro-ph/9810509 [astro-ph]].
\bibitem{Dimopoulos:2019gpz}
K.~Dimopoulos and L.~Donaldson-Wood,
Phys. Lett. B \textbf{796}, 26-31 (2019)
[arXiv:1906.09648 [gr-qc]].
\bibitem{Lima:2019yyv}
G.~B.~F.~Lima and R.~O.~Ramos,
Phys. Rev. D \textbf{100}, no.12, 123529 (2019)
[arXiv:1910.05185 [astro-ph.CO]].
\bibitem{Ford:1986sy}
L.~H.~Ford,
Phys. Rev. D \textbf{35}, 2955 (1987).

\bibitem{Chun:2009yu}
E.~J.~Chun, S.~Scopel and I.~Zaballa,
JCAP \textbf{07}, 022 (2009)

[arXiv:0904.0675 [hep-ph]].
\bibitem{Felder:1998vq}
G.~N.~Felder, L.~Kofman and A.~D.~Linde,
Phys. Rev. D \textbf{59}, 123523 (1999)
[arXiv:hep-ph/9812289 [hep-ph]].
\bibitem{Campos:2002yk}
A.~H.~Campos, H.~C.~Reis and R.~Rosenfeld,
Phys. Lett. B \textbf{575}, 151-156 (2003)
[arXiv:hep-ph/0210152 [hep-ph]].

\bibitem{Feng:2002nb}
B.~Feng and M.~z.~Li,
Phys. Lett. B \textbf{564}, 169-174 (2003)
[arXiv:hep-ph/0212213 [hep-ph]].

\bibitem{BuenoSanchez:2007jxm}
J.~C.~Bueno Sanchez and K.~Dimopoulos,
JCAP \textbf{11}, 007 (2007)
[arXiv:0707.3967 [hep-ph]].

\bibitem{Dimopoulos:2018wfg}
K.~Dimopoulos and T.~Markkanen,
JCAP \textbf{06}, 021 (2018)
[arXiv:1803.07399 [gr-qc]].


\bibitem{Opferkuch:2019zbd}
T.~Opferkuch, P.~Schwaller and B.~A.~Stefanek,
JCAP \textbf{07}, 016 (2019)
[arXiv:1905.06823 [gr-qc]].

\bibitem{Bettoni:2021zhq}
D.~Bettoni, A.~Lopez-Eiguren and J.~Rubio,
[arXiv:2107.09671 [hep-ph]].


\bibitem{Sahni:2001qp}
V.~Sahni, M.~Sami and T.~Souradeep,
Phys. Rev. D \textbf{65}, 023518 (2002)
[arXiv:gr-qc/0105121 [gr-qc]].




\bibitem{Berera:2008ar}
A.~Berera, I.~G.~Moss and R.~O.~Ramos,
Rept. Prog. Phys. \textbf{72}, 026901 (2009)
[arXiv:0808.1855 [hep-ph]].
\bibitem{Bastero-Gil:2019gao}
M.~Bastero-Gil, A.~Berera, R.~O.~Ramos and J.~G.~Rosa,
Phys. Lett. B \textbf{813}, 136055 (2021)
[arXiv:1907.13410 [hep-ph]].
\bibitem{Bastero-Gil:2010dgy}
M.~Bastero-Gil, A.~Berera and R.~O.~Ramos,
JCAP \textbf{09}, 033 (2011)
[arXiv:1008.1929 [hep-ph]].
\bibitem{Bastero-Gil:2012akf}
M.~Bastero-Gil, A.~Berera, R.~O.~Ramos and J.~G.~Rosa,
JCAP \textbf{01}, 016 (2013)
[arXiv:1207.0445 [hep-ph]].
\bibitem{Bastero-Gil:2016qru}
M.~Bastero-Gil, A.~Berera, R.~O.~Ramos and J.~G.~Rosa,
Phys. Rev. Lett. \textbf{117}, no.15, 151301 (2016)
[arXiv:1604.08838 [hep-ph]].

\bibitem{Geng:2015fla}
C.~Q.~Geng, M.~W.~Hossain, R.~Myrzakulov, M.~Sami and E.~N.~Saridakis,
Phys. Rev. D \textbf{92}, no.2, 023522 (2015)
[arXiv:1502.03597 [gr-qc]].
\bibitem{Geng:2017mic}
C.~Q.~Geng, C.~C.~Lee, M.~Sami, E.~N.~Saridakis and A.~A.~Starobinsky,
JCAP \textbf{06}, 011 (2017)
[arXiv:1705.01329 [gr-qc]].


\bibitem{Ahmad:2019jbm}
S.~Ahmad, A.~De Felice, N.~Jaman, S.~Kuroyanagi and M.~Sami,
Phys. Rev. D \textbf{100}, no.10, 103525 (2019)
[arXiv:1908.03742 [gr-qc]].
\bibitem{AresteSalo:2020yxl}
L.~Areste Salo and J.~Haro,
Eur. Phys. J. C \textbf{81}, no.2, 105 (2021)
[arXiv:2009.12912 [gr-qc]].

\bibitem{Rezazadeh:2015dia}
K.~Rezazadeh, K.~Karami and S.~Hashemi,
Phys. Rev. D \textbf{95}, no.10, 103506 (2017)
[arXiv:1508.04760 [gr-qc]].




\bibitem{DeSimone:2016ofp}
A.~De Simone and T.~Kobayashi,
JCAP \textbf{08}, 052 (2016)
[arXiv:1605.00670 [hep-ph]].


\bibitem{DeFelice:2002ir}
A.~De Felice, S.~Nasri and M.~Trodden,
Phys. Rev. D \textbf{67}, 043509 (2003)
[arXiv:hep-ph/0207211 [hep-ph]].
\bibitem{Arbuzova:2016spc}
E.~V.~Arbuzova, A.~D.~Dolgov and V.~A.~Novikov,
Phys. Rev. D \textbf{94}, no.12, 123501 (2016)
[arXiv:1607.01247 [astro-ph.CO]].
\bibitem{Dasgupta:2018eha}
A.~Dasgupta, R.~K.~Jain and R.~Rangarajan,
Phys. Rev. D \textbf{98}, no.8, 083527 (2018)
[arXiv:1808.04027 [hep-ph]].
\bibitem{Kolb:1990vq}
E.~W.~Kolb and M.~S.~Turner,
Front. Phys. \textbf{69}, 1-547 (1990).








\bibitem{suratna}
S.~Das,
Phys. Rev. D \textbf{99}, no.6, 063514 (2019)
[arXiv:1810.05038 [hep-th]].

\bibitem{Copeland:1997et}
E.~J.~Copeland, A.~R.~Liddle and D.~Wands,
Phys. Rev. D \textbf{57}, 4686-4690 (1998)
[arXiv:gr-qc/9711068 [gr-qc]].

\bibitem{Copeland:2006wr}
E.~J.~Copeland, M.~Sami and S.~Tsujikawa,
Int. J. Mod. Phys. D \textbf{15}, 1753-1936 (2006)
[arXiv:hep-th/0603057 [hep-th]].
\bibitem{Haro:2019peq}
J.~Haro, J.~Amor\'os and S.~Pan,
Eur. Phys. J. C \textbf{80}, no.5, 404 (2020)
[arXiv:1908.01516 [gr-qc]].
\bibitem{Steinhardt:1999nw}
P.~J.~Steinhardt, L.~M.~Wang and I.~Zlatev,
Phys. Rev. D \textbf{59}, 123504 (1999)
[arXiv:astro-ph/9812313 [astro-ph]].

\bibitem{Skugoreva:2019blk}
M.~A.~Skugoreva, M.~Sami and N.~Jaman,
Phys. Rev. D \textbf{100}, no.4, 043512 (2019)
[arXiv:1901.06036 [gr-qc]].
\bibitem{Barreiro:1999zs}
T.~Barreiro, E.~J.~Copeland and N.~J.~Nunes,
Phys. Rev. D \textbf{61}, 127301 (2000)
[arXiv:astro-ph/9910214 [astro-ph]].


\bibitem{Amendola:1999er}
L.~Amendola,
Phys. Rev. D \textbf{62}, 043511 (2000)
doi:10.1103/PhysRevD.62.043511
[arXiv:astro-ph/9908023 [astro-ph]].
\bibitem{Gumjudpai:2005ry}
B.~Gumjudpai, T.~Naskar, M.~Sami and S.~Tsujikawa,
JCAP \textbf{06}, 007 (2005)
[arXiv:hep-th/0502191 [hep-th]].

\bibitem{Wetterich:2013jsa}
C.~Wetterich,
Phys. Rev. D \textbf{89}, no.2, 024005 (2014)
doi:10.1103/PhysRevD.89.024005
[arXiv:1308.1019 [astro-ph.CO]].
\bibitem{Hossain:2014xha}
M.~W.~Hossain, R.~Myrzakulov, M.~Sami and E.~N.~Saridakis,
Phys. Rev. D \textbf{90}, no.2, 023512 (2014)
[arXiv:1402.6661 [gr-qc]].

\bibitem{DeFelice:2010aj}
A.~De Felice and S.~Tsujikawa,
Living Rev. Rel. \textbf{13}, 3 (2010)
[arXiv:1002.4928 [gr-qc]].
\bibitem{samrev}
M.~Sami and R.~Gannouji,
[arXiv:2106.00843 [gr-qc]].
\bibitem{Sami:2019hbq}
M.~Sami, S.~Myrzakul and M.~Al Ajmi,
Phys. Dark Univ. \textbf{30}, 100675 (2020)
[arXiv:1912.12026 [gr-qc]].

\bibitem{Sami:2021ufn}
M.~Sami and R.~Gannouji,

[arXiv:2106.00843 [gr-qc]].

\bibitem{Bastero-Gil:2011clw}
M.~Bastero-Gil, A.~Berera, R.~O.~Ramos and J.~G.~Rosa,
Phys. Lett. B \textbf{712}, 425-429 (2012)
[arXiv:1110.3971 [hep-ph]].

\bibitem{Figueroa:2018twl}
D.~G.~Figueroa and E.~H.~Tanin,
JCAP \textbf{10}, 050 (2019)
[arXiv:1811.04093 [astro-ph.CO]].
\bibitem{Kuroyanagi:2008ye}
S.~Kuroyanagi, T.~Chiba and N.~Sugiyama,
Phys. Rev. D \textbf{79}, 103501 (2009)
[arXiv:0804.3249 [astro-ph]].
\bibitem{Boyle:2005se}
L.~A.~Boyle and P.~J.~Steinhardt,
Phys. Rev. D \textbf{77}, 063504 (2008)
[arXiv:astro-ph/0512014 [astro-ph]].
\bibitem{Watanabe:2006qe}
Y.~Watanabe and E.~Komatsu,
Phys. Rev. D \textbf{73}, 123515 (2006)
[arXiv:astro-ph/0604176 [astro-ph]].
\bibitem{Moore:2014lga}
C.~J.~Moore, R.~H.~Cole and C.~P.~L.~Berry,
Class. Quant. Grav. \textbf{32}, no.1, 015014 (2015)
[arXiv:1408.0740 [gr-qc]].

\bibitem{mrgsarma}
M.~R.~Gangopadhyay, S.~Myrzakul, M.~Sami and M.~K.~Sharma,
Phys. Rev. D \textbf{103}, no.4, 043505 (2021)
[arXiv:2011.09155 [astro-ph.CO]].
\bibitem{mbggw}
M.~Bastero-Gil, M.~S.~Díaz-Blanco, arXiv: 2105.08045 [hep-ph].
\end{thebibliography}
\end{document}